\documentstyle[12pt]{article}
\newcommand{\newc}{\newcommand}    
\newc{\ra}{\rightarrow} 
\newc{\lra}{\leftrightarrow} 
\newc{\beq}{\begin{equation}} 
\newc{\eeq}{\end{equation}} 
\newc{\barr}{\begin{eqnarray}} 
\newc{\earr}{\end{eqnarray}} 
\newc{\texa}{\textstyle}
\newc{\paral}{\parallel}
\newc{\und}{\underline}
\newc{\pars}{\partial}
\newc{\nonu}{\nonumber \\}

\begin{document}
\thispagestyle{empty}
\begin{center}
{\Large \bf Ideal magnetohydrodynamic equilibria\\ \vspace{3mm}
	   with helical symmetry and incompressible flows \vspace{5mm} }\\
\large\bf
{\large G. N. Throumoulopoulos\footnote                           
{Permanent  address: Section of Theoretical Physics,
 Physics Department, University of Ioannina 
 GR  451 10 Ioannina, Greece} and H. Tasso \vspace{1mm}\\
{\it   Max-Planck-Institut f\"{u}r Plasmaphysik, EURATOM
Association \\ \vspace{1mm}
 D-85748 Garching, Germany } }
\end{center}
\vspace{2mm}
\vspace{0.5cm}
%
\begin{center}
{\large\bf Abstract} 
\end{center}
 A recent study on axisymmetric ideal magnetohydrodynamic equilibria 
 with incompressible flows [H. Tasso and G. N. Throumoulopoulos, Phys.
 Plasmas {\bf 5}, 2378 (1998)] is extended to the  generic case of
 helically symmetric equilibria with incompressible flows.
It is shown that  the  equilibrium states of the system under consideration
are governed by
 an elliptic partial differential equation for the helical 
magnetic flux function $\psi$ 
containing
five surface quantities 
along with a relation for the pressure. 
The above mentioned 
 equation can  
be transformed to one possessing differential part identical in form to
the corresponding static equilibrium equation, which is amenable to
several classes of analytic solutions. 
In particular, equilibria with 
electric fields perpendicular to the magnetic surfaces 
and non-constant-Mach-number flows are constructed. Unlike 
the case in axisymmetric equilibria with isothermal magnetic surfaces,
helically symmetric $T=T(\psi)$ equilibria are over-determined, i.e., in this 
case the equilibrium equations
reduce to a set of eight ordinary differential equations with seven surface
quantities.
In addition, it is proved the non-existence of incompressible helically
symmetric
equilibria with 
 (a)   purely helical flows (b) 
non-parallel flows with isothermal magnetic surfaces  and 
the magnetic field modulus being a surface quantity
(omnigenous equilibria).  
\newpage
\setcounter{page}{1}

\begin{center} 
{\large \bf I.\ \ Introduction}
\end{center}            
Understanding the equilibrium  properties of  magnetically
confined plasmas is  one of the  basic  objectives of
fusion research (see citations of Ref. \cite{TaTh98}).    
In particular, in the last fifteen years,  
equilibria with  flows which are induced
when  
either neutral beams 
or/and   electromagnetic power are employed  to heat 
the plasma  
of magnetic confinement systems, has become an issue  
of increasing interest.  
These flows 
are  usually associated with radial electric fields which play a 
role in the transitions to the improved confinement regimes 
\cite{Bu,ErWa,BaKi,ShRu}.

In a series of  papers \cite{TaTh98,TaTh98a,ThTa97,ThPa96}
we  investigated the magnetohydrodynamic 
(MHD) equilibrium of ideal plasmas with   incompressible flows and translational
as well as axial symmetry.  The main conclusions of the above mentioned 
studies are the following:
(a)  If the equilibrium flows of   cylindrical plasmas 
with arbitrary cross-sectional shapes
are purely poloidal,   they should be incompressible.
(b) Exact equilibria were constructed for constant poloidal flow Mach
numbers $M^2$ (see definition (\ref{21}) in Sec. II)  
for both cylindrical and axisymmetric configurations.
(c) For the physically appealing class of cylindrical equilibria  
with isothermal magnetic surfaces  their
 cross-sections  must be circular at large, 
while no restriction is imposed on the magnetic surfaces of 
axisymmetric $T=T(\psi)$ equilibria, apart from the vicinity of the  
magnetic axis where they should be circular.

The aim of the present report is to extend our previous studies 
to the generic case of helically symmetric ideal MHD equilibria
with incompressible flows which  describe the steady states of the
plasmas of straight stellarators.    
It is shown that  the  equilibrium states 
satisfy 
 an elliptic partial differential equation for the helical 
magnetic flux function $\psi$ 
containing
five surface quantities 
along with a relation for the pressure. 
This is the object of Sec. II.
In Sec. III the above mentioned 
differential equation is transformed
to one possessing differential part identical in form to
the corresponding static equilibrium equation, which permits the
derivation of
several classes of analytic solutions 
with {\em non-constant} $M^2$ and   
 differentially varying  
electric fields perpendicular to the magnetic surfaces.
We obtain exact solutions 
with flows in the symmetry direction,  parallel 
to the magnetic field and in arbitrary direction. 
Equilibria with isothermal magnetic surfaces are then examined
in Sec. IV. Sec. V summarizes our conclusions.
\newpage
\begin{center}
{\large\bf II.\ \ Equilibrium equations} 
\end{center}

The ideal MHD equilibrium states of  plasma
 flows are governed by the following set of
equations, written in standard notations and convenient units:
\begin{equation}
{\bf\nabla} \cdot (\rho {\bf v}) = 0 
					    \label{1}
\end{equation}
\begin{equation}
\rho ({\bf v} \cdot {\bf\nabla})  {\bf v} = {\bf j}
\times {\bf B} - {\bf\nabla} P 
					    \label{2}
\end{equation}
\begin{equation}
{\bf\nabla} \times  {\bf E} = 0 
					    \label{3}
\end{equation}
\begin{equation}
{\bf\nabla}\times {\bf B} = {\bf j }
					    \label{4}
\end{equation}
\begin{equation}
{\bf\nabla} \cdot {\bf B} = 0 
					    \label{5}
\end{equation}
\begin{equation}
{\bf E} +{\bf v} \times {\bf B} = 0.
					    \label{6}
\end{equation}
The system under consideration is a helically symmetric 
magnetically confined plasma with flow. To describe this configuration
we introduce cylindrical coordinates $R, z, \phi$ with $z$ along the 
rotation axis; helical symmetry implies that any physical 
quantity depends only on $R$ and $u=l\phi + k z$. Though a helix is 
characterized by just one parameter, for instance $a=k/l$, it is convenient to
keep both $l$ and $k$ in order to discuss the two limiting cases $l=0$ 
corresponding to axisymmetry and $k=0$ corresponding to translational 
symmetry. Also, the vector 
\beq
{\bf h} = \frac{l \nabla z- k R^2\nabla \phi}{l^2 + k^2 R^2}
						     \label{7}
\eeq
is introduced which is tangent to the helix $r=$ const. $u=$ const., with  
$\nabla u=l\nabla \phi + k \nabla z$. It then follows from Eq. (\ref{7})
that ${\bf h}\cdot \nabla G(R,u) =0$ for any function $G(R,u)$.
Also, ${\bf h}$ has the properties
\beq
\nabla\cdot {\bf h}  = 0, 
\ \ \ \nabla\times {\bf h} = \frac{-2kl}{l^2 + k^2 R^2}{\bf h}.
						      \label{7a}
\eeq
The divergence-free fields,
i.e. the magnetic field $\bf B$, the current 
density ${\bf j}$ and the mass flow $\rho{\bf v}$ can be
expressed in terms of the stream functions $\psi(R,u)$, 
$I(R,u)$, $F(R,u)$ and $\Theta(R,u)$ as
\begin{equation} 
{\bf B} = I {\bf h} +
 {\bf h} \times \nabla\psi,
					     \label{8}
 \end{equation}
\begin{equation}
{\bf j} = ({\cal L}\psi - 2kl h^2 I){\bf h}  - {\bf h} \times \nabla I
					     \label{9}
\end{equation}
and
\begin{equation}
\rho {\bf v} =\Theta {\bf h} +  {\bf h} \times{\nabla} F,
					       \label{10}
\end{equation}
where ${\cal L}$ is an elliptic operator defined by
\begin{eqnarray}
{\cal L \psi}&\equiv & \frac{1}{h^2}\nabla\cdot(h^2 \nabla \psi)=                                       
	   \frac{1}{Rh^2}\left[\frac{\pars}{\pars R}\left(Rh^2 
	   \frac{\pars}{\pars R}\right) + 
	   \frac{1}{R} \frac{\pars^2}{\pars u^2}\right]\psi \nonumber \\
	&=&\left[\frac{\pars^2}{\pars R^2} +\frac{1}{R}
	  \left(\frac{l^2 - k^2R^2}{l^2 +k^2R^2}\right)\frac{\pars}{\pars R}
	  + \frac{l^2 + k^2R^2}{R^2}\frac{\pars^2}{\pars u^2} \right]\psi,
						    \label{11}
\end{eqnarray} 
and $h^2\equiv {\bf h}\cdot{\bf h}=1/(l^2 + k^2R^2) $.     
Eqs. (\ref{1})-(\ref{6}) can be reduced  by means of 
certain integrals of the system, which are shown to be surface quantities.
To identify three of these quantities,  the time independent  electric field
is expressed by ${\bf E} = - {\bf \nabla} \Phi$ and the Ohm's law
(\ref{6}) is projected along ${\bf h}$,  $\bf B$ and $\nabla \psi$, 
respectively, yielding  $F=F(\psi)$, $\Phi=\Phi(\psi)$ and
\begin{equation}
\frac{h^2}{\rho }(IF^\prime-\Theta)= \Phi^\prime.
					      \label{12}
\end{equation}
(The prime denotes differentiation with respect to $\psi$).  A fourth 
surface quantity is derived from the component of momentum conservation
equation (\ref{2}) along ${\bf h}$: 
\begin{equation}
I\left(1-\frac{(F^\prime)^2}{\rho}\right)
+\frac{F^\prime \Phi^\prime}{h^2} \equiv  X(\psi).
					      \label{13}
\end{equation}
 From Eq. (\ref{13}) it follows that,
unlike the case in static equilibria, 
					     $I$                
is not a surface quantity. This implies that in flowing
plasmas the  current density 
does not lie on magnetic surfaces [see Eq. (\ref{9})]. 
Solving Eqs. (\ref{12}) and (\ref{13})
for $I$ and $\Theta$ one obtains
\beq
I=\frac{X -F^\prime\Phi^\prime/h^2}{1-(F^\prime)^2/\rho}
					       \label{13a}
\eeq
and 
\beq
\Theta = \frac{F^\prime X -\rho\Phi^\prime/h^2}{1-(F^\prime)^2/\rho}. 
						 \label{13b}
\eeq
Also, on account of Eq. (\ref{12}), the  velocity [Eq. (\ref{10})]
can be written in the form 
\begin{equation}
{\bf v} = \frac{F^\prime}{\rho} {\bf B} - \frac{\Phi^\prime}{h^2} 
	  {\bf h}.
					      \label{13c}
\end{equation}
With the aid of 
Eqs. (\ref{10})-(\ref{13c}), the components of 
Eq. (\ref{2}) along $\bf B$ and  perpendicular to a
magnetic surface are put in the respective forms
\begin{equation}
{\bf B} \cdot \left\lbrack{\bf \nabla} \left(\frac{v^2}{2}
+ \frac{\Theta}{\rho}\Phi^\prime\right)
+ \frac{\nabla P}{\rho}\right\rbrack = 0 
					       \label{14}
 \end{equation}
and
\begin{eqnarray} 
\left\{ {\bf \nabla} \cdot 
\left\lbrack\left(1- \frac{(F^\prime)^2}{\rho}\right)
h^2{\bf \nabla}\psi   \right\rbrack 
+ \frac{F^\prime F^{\prime\prime} }{\rho}{h^2|\nabla\psi|^2}
-2klh^4 X \right\}
   |\nabla \psi|^2
& & \nonumber \\ 
+ \left\lbrack\frac{\rho}{2}\left(\nabla v^2
		- h^2 {\nabla (\Theta/ \rho)^2}\right)
   + h^2 \frac{\nabla I^2}{2 } +\nabla P  \right\rbrack\cdot \nabla\psi = 0 & &
						\label{15} 
\end{eqnarray}
It is pointed out that  Eqs. (\ref{14}) and (\ref{15})
are valid for any equation of state for the plasma. 

In order
to reduce the equilibrium equations further, we employ the
incompressibility condition            
\begin{equation}
\nabla\cdot {\bf v} = 0,
					    \label{16}
\end{equation}
which on account of $\nabla\cdot \rho {\bf v}=0$ implies that
the density is a surface quantity, 
\begin{equation}
\rho=\rho(\psi).
					    \label{17}   
\end{equation}
Consequently,  from Eqs. (\ref{12}) and (\ref{13})  it follows that,
similar to axisymmetric incompressible equilibria with purely poloidal flows  
\cite{TaTh98}, 
helically symmetric plasmas with purely helical flows ($\Theta=0$)
can not exist;  the only possible equilibria of this kind are  cylindrical
configurations with arbitrary cross sectional shapes \cite{ThPa96}.
With the aid of Eq. (\ref{17}),  Eq. (\ref{14})
can be integrated to yield an expression for the pressure, i.e.
\begin{equation} 
P = P_s
(\psi) - \rho \left(\frac{v^2}{2}
	       + \frac{\Phi^\prime\Theta}{\rho} \right).
						\label{19}    
\end{equation}
We note here that, unlike in static 
equilibria, in the presence of flow  magnetic surfaces in general
do not coincide with
isobaric surfaces  because Eq. (\ref{2})
implies that ${\bf B} \cdot {\bf \nabla} P$ in
general differs from zero.
In this respect, the term $P_s(\psi)$ is the static part of 
the pressure which
does not vanish  when ${\bf v} = {\bf 0}$.  

Eq. (\ref{15}) has  a singularity when  
\begin{equation}
\frac{\left(F^\prime \right)^2}{\rho}=1. 
						  \label{20}
\end{equation}
On the basis of  the
definitions   
$v_{h}^2\equiv 
\frac{\textstyle (F^\prime)^2 h^2|\nabla \psi|^2}{\textstyle \rho}$
for the helical velocity component, 
$v_{Ah}^2\equiv\frac{\textstyle  h^2 |\nabla \psi|^2}
{\textstyle \rho}$ 
for the Alfv\'en velocity associated with the
helical magnetic field, and the Mach number
\begin{equation}
 M^2\equiv\frac{v_h^2}{v_{Ah}^2}= \frac{(F^\prime )^2}{\rho},
						   \label{21} 
\end{equation}
Eq. (\ref{20}) can be written as
$
M^2= 1.
$
If it is now assumed that
$\frac{\textstyle (F^\prime)^2}{\textstyle \rho}\neq 1$
and Eq. (\ref{19}) is inserted into Eq. (\ref{15}), 
the latter reduces to
the {\em elliptic} differential equation
\begin{eqnarray} 
(1-M^2) {\cal L} \psi - 
	     \frac{1}{2}(M^2)^\prime |\nabla \psi|^2 
	     -2klh^2X
					   & & \nonumber \\ 
+ \frac{1}{2}\left(\frac{X^2}{1-M^2}\right)^\prime
+ \frac{1}{h^2}\left(P_s -  \frac{X F^\prime\Phi^\prime}{1-M^2}\right)^\prime 
+ \frac{1}{2 h^4 }\left(\frac{\rho (\Phi^\prime)^2}{1-M^2}\right)^\prime
    = 0.& & 
						    \label{21a}
\end{eqnarray}
This is the equilibrium equation for a helically symmetric plasma
with incompressible flows.
 Eq. (\ref{21a}) contains the arbitrary
surface quantities $F(\psi)$, $\Phi(\psi)$, $X(\psi)$,
$\rho(\psi)$ and $P_s(\psi)$ which must be found from other
physical considerations. 
Inspection of Eqs. (\ref{21a}), (\ref{9}) and (\ref{13a}) shows  
that the singularity $M^2=1$ 
is the limit at which the confinement 
can be assured by the toroidal current density ${\cal L}{\bf h}$ alone.
For $M^2\neq 1$ the derivative of $X^2/(1-M^2)$, related to the derivative
of the magnetic field component in the symmetry direction 
by Eq. (\ref{13}),                                  
partly compensates for the pressure gradient and inertial flow forces. 

It should be noted here  
that the equilibrium of
helically symmetric equilibria with incompressible flows 
was investigated in Refs. \cite{Ts,ViTs,ViFe} and \cite{PaPl},  
where 
 a set   of
an elliptic differential equation and a nonlinear 
algebraic equation 
was obtained  
(e. g. Eqs.   (38) and (40) 
of Ref. \cite{PaPl}). 
Those equations,  
however, are coupled 
(and consequently one has to solve the one 
equation and then examine {\em a posteriori} whether the solutions are 
compatible  with the other),
a property  which makes the derivation of  
analytic  solutions  tedious.  
On the other side, 
as concerns the present  study, one has to solve  
  a {\em single} differential equation [Eq.  (\ref{21a})]  
which is decoupled  from Eq. (\ref{19}). Once the  solutions of 
Eq. (\ref{21a}) are known, Eq. (\ref{19})  determines only the pressure. 
Moreover, as shown in section III, Eq. (\ref{21a}) can be transformed 
to one  which is amenable to several classes of analytic solutions
describing equilibria with non-constant-poloidal-flow Mach numbers 
[$(d/dU)M^2\neq 0)$].

\begin{center}
{\large\bf III.\ \ Analytic equilibrium solutions}
\end{center}

Eq. (\ref{21a})  under the transformation \cite{Mo,Cl}
\beq
U(\psi)= \int_0^{\psi}\, [1-M^2(\psi^{\prime})]^{1/2} \,
 d\psi^{\prime},\ \    M^2<1
						       \label{22}
\eeq
reduces (after dividing by $(1-M^2)^{1/2}$) to 
\begin{eqnarray} 
 {\cal L} U  
 -2klh^2\frac{X}{(1-M^2)^{1/2}}
 + \frac{1}{2}\frac{d}{d U} \left(\frac{X^2}{1-M^2}\right)
+ \frac{1}{h^2}\frac{d}{d U}\left(P_s -  X \frac{dF}{d U}\frac{d\Phi}{d U} \right) 
						& &  \nonumber  \\
+ \frac{1}{2h^4 }\frac{d}{d U}
  \left[\rho \left( \frac{d\Phi}{d U}\right)^2\right]
    = 0.& &                                      \label{23}
\end{eqnarray}
It is noted here that the requirement $M^2<1$ in transformation (\ref{22})
implies that $v_{h}^2<v_s^2$, where $v_s= (\gamma P/\rho)^{1/2}$ is the
sound speed. This follows from 
Eqs. (\ref{21}) and (in Gaussian units) 
$$
\left(\frac{v_s}{v_{Ah}}\right)^2 = 
(\gamma/2)\frac{8\pi P}{h^2|\nabla \psi|^2}\approx 1.
$$ 
Since, 
according to experimental evidence in stellarators \cite{ErWa,BaKi,ShRu}, 
the (maximum) value
of the ion poloidal velocity (corresponding to $v_{h}$) in the edge region 
during the L-H transition
is of the order of 1 Km/sec and the ion temperature is of the order 
of $100$ eV, the scaling $v_h \ll v_s$ is satisfied in this region.
Therefore, 
the restriction $M^2<1$ is of non-operational relevance.
Also, solving   Eq. (\ref{23}) only is sufficient 
to determine the equilibrium  because it is  not $\psi(U)$ itself 
but $\nabla \psi(U) =(1-M^2(U))^{1/2}\nabla U$ which is needed to
obtain any equilibrium quantity.

Three classes of exact equilibria of Eq. (\ref{23}) 
can be constructed as follows.

\begin{center}
{\large\it (a) Flows in the symmetry direction}
\end{center}
This kind of equilibria correspond to  $F^\prime=0$.
 From Eq. (\ref{13a}) it then follows that the function
 $I$ becomes a surface quantity, $I=X/(1-M^2)$, 
and Eq. (\ref{23})  reduces to
\begin{eqnarray} 
 {\cal L} U  
 -2klh^2\frac{X}{(1-M^2)^{1/2}}
 + \frac{1}{2}\frac{d}{d U} \left(\frac{X^2}{1-M^2}\right)
+ \frac{1}{h^2}\frac{d P_s}{d U} 
						& & \nonumber \\ 
+ \frac{1}{2 h^4}\frac{d}{d U}
  \left[\rho \left( \frac{d\Phi}{d U}\right)^2\right]
    = 0.& & 
						    \label{24}
\end{eqnarray}
Eq. (\ref{24}) can be linearized and  solved analytically 
for several choices of the surface quantities it contains. For example,
two classes of polynomial solutions with respect to 
$R$ and $u$ can be constructed for 
$(d/dU)[X(1- M^2)^{-1/2}] =$  constant, $(d/dU)P_s=$ constant and (i) 
$\rho (d\Phi/d U)^2=$ const. and (ii) $\rho (d\Phi/d U)^2\propto U$. 
In case (ii), owing to the flow term containing the factor $1/h^4$ in Eq. (\ref{23}), the polynomials 
can describe configurations with either a single magnetic axis
or two magnetic axes. 
A simple equilibrium  with  single magnetic axis corresponding  
to $(d/dU)[X(1- M^2)^{-1/2}] =0$ is given by  
\begin{equation}
 U=\frac{U_c}{5 R_c^6}R^2
      (9R_c^4 - 3R_c^2R^2 - R^4 -\delta R^2 u^2),
						       \label{25}
\end{equation}
where $U_c$ is the value of the flux function at the position
of the magnetic axis ($u=0$, $R=R_c$) and $\delta$ is a parameter
related to  the shape of the flux surfaces.

\begin{center}
{\large\it (b) Flows parallel to $\bf B$}
\end{center}

Equilibria with $\bf B$-aligned flows
correspond to $\Phi^\prime=0$.
Eq. (\ref{23}) then reduces to
\beq
 {\cal L} U  -2klh^2\frac{X}{(1-M^2)^{1/2}}
 + \frac{1}{2}\frac{d}{d U} \left(\frac{X^2}{1-M^2}\right)
+ \frac{1}{h^2}\frac{d P_s}{d U}  =0,
						\label{26}
\eeq
which is similar in form to the equation governing
static equilibria. (Analytic static helically symmetric equilibria
have been derived in Refs. \cite{Ka,CoLo,Co}). Eq. (\ref{26})
can be  linearized and solved for (a) 
$(d/dU)[X(1- M^2)^{-1/2}] = c_0 + c_1 U$ and  $(d/dU)P_s= d_0 + d_1 U$, 
where $c_0$, $c_1$, $d_0$ and $d_1$ are constant quantities.
The simplest solution corresponding to $c_0=c_1=d_1=0$   
 is given by 
\beq
U = U_c\frac{R^2}{R_c^4}(2R_c^2 - R^2 -\delta u^2).
						  \label{27}
\eeq
For $l=0$ and $k=1$ Eq. (\ref{27}) reduces to the Hill's 
vortex axisymmetric equilibrium \cite{Tho}.

\begin{center}
{\large\it (c) Non-parallel flows} 
\end{center}

Equilibria of this kind are  of particular interest because
non-parallel flows with  non-vanishing poloidal (helical) components  
are associated with radial electric fields 
which play a role in the transitions to  
improved confinement regimes in tokamaks \cite{Bu} and stellarators
\cite {BaKi}. Analytic,  smooth equilibrium solutions of Eq. (\ref{23})
can be derived for (i) $\rho (d\Phi/d U)^2=$ constant 
and  (ii) $\rho (d\Phi/d U)^2=c_2 U,\ c_2=\mbox{constant}$. 
Because of the departure of the isobaric surfaces and current density surfaces
from the magnetic surfaces,  however, 
in  case (i) it is not possible to obtain physically  
plausible  equilibria,  viz.
equilibria with  singly peaked
density and pressure profiles (e. g. on the plane $u=0$) vanishing on the 
plasma surface,  a current density profile  not possessing  
a component along $\nabla U$ on the plasma surface and finite $E$ 
throughout the helical  cross-section. In  case (ii)
a class of physically well behaved equilibria is derived with 
the ansatz
\begin{eqnarray*}
\frac{X}{(1-M^2)^{1/2}}= c_0=\mbox{const},\ \ 
P_s -X\frac{dF}{d U}\frac{d\Phi}{d U}\propto U, & &  \\ 
M^2= c_1 U^\mu\ (c_1 = \mbox{const.},\ \mu>1) 
\ \  \mbox{and}\ \  \rho\propto U^\lambda \ (0<\lambda<1).   & &
\end{eqnarray*}
Equation (\ref{23}) then becomes
\beq
 {\cal L} U  -2klh^2c_0 + \frac{c_1}{h^2} + \frac{c_2}{2h^4}.
						   \label{28}
\eeq
A particular solution of Eq. (\ref{28}) is given by Eq. (\ref{25}).  
The radial electric field  on the plane $u=0$ following from                                      
Eq. (\ref{25}) is given by
\beq
{\bf E}_R =-\left.\frac{d\Phi}{d U} \nabla U\right|_{u=0}
	  =-U^{(1-\lambda)/2}\frac{\pars U}{\pars R}\nabla R.
						      \label{28a}
\eeq
Eq. (\ref{28a}) implies that $|{\bf E}_R|$ 
vanishes 
at the magnetic axis $R=Rc$ and on the plasma surface 
$R=R_s=[3(\sqrt{5} -1)/2]^{1/2}]R_c$ (which is determined  by $U(u=0, R)=0$) 
and has an extremum 
in between  
$R_c$ and $R_s$, the position of which is a
function of the parameters  $R_c$ and $\lambda$.   This 
profile is consistent with those observed during the L-H transition
in stellarators \cite{ErWa,BaKi,ShRu}.
\begin{center}
{\large\bf III.\ \ Equilibria with isothermal magnetic surfaces}
\end{center}

For fusion plasmas the thermal conduction along $\bf B$ is fast 
compared to  the heat transport perpendicular
to a magnetic surface and therefore equilibria
with isothermal magnetic surfaces are of particular interest.
It is recalled  that for cylindrical plasmas
the relation $T=T(\psi)$ restricts the cross sections of the
magnetic surfaces to be circular 
at large, while for axisymmetric plasmas 
no restriction is imposed on the shape of the magnetic surfaces apart
from the vicinity of the magnetic axis where their cross sections should
be circular. 

Under the assumption that the plasma
obeys the ideal gas low
$P = \hat{R} \rho T$, Eqs. (\ref{19}), (\ref{12}) and (\ref{13c}) lead
to the following expression for the magnetic field modulus:
\begin{equation}
  |{\bf B}|^2 = \Xi(\psi) + \frac{H(\psi)}{h^2},
							 \label{30a}
\end{equation}
where $\Xi(\psi) \equiv 2(P_s -P)\rho/(F^\prime)^2$ and
$H(\psi) \equiv (\rho\Phi^\prime/F^\prime)^2$.
Consequently, apart from the case of field
aligned flows ($H=0$), omnigenous  equilibria,
viz. equilibria with $|\bf B|$ being a surface quantity, are not
possible. It is  noted that this property  also holds for axisymmetric
equilibria \cite{TaTh98}. 

Solving the set of equations (\ref{21a}) and (\ref{30a}) for 
$|\nabla \psi|^2$ and ${\cal L} \psi$
one obtains
\begin{equation}
 \left| \nabla \psi\right|^2 =
 \left(\frac{\partial \psi}{\partial R} \right)^2
+  \frac{1}{R^2 h^2}\left(\frac{\partial \psi}{\partial u} \right)^2
 = 2\left(i(\psi) + \frac{j(\psi)}{h^2} +  \frac{k(\psi)}{h^4}\right)
							 \label{31}
\end{equation}
and
\begin{equation}
{\cal L}\psi= -h^2 w(\psi) - f(\psi) 
-\frac{g(\psi)}{h^2} - \frac{d(\psi)}{h^4},
							 \label{32}
\end{equation}
where $i(\psi)$, $j(\psi)$, $k(\psi)$, $w(\psi)$, $f(\psi)$, $g(\psi)$  
and $d(\psi)$ are known functions of $\psi$.
With the  introduction of  the quantities  $x\equiv R^2$,
  $p=\partial \psi/\partial x$,
$q=\partial \psi/\partial u$, $r=\partial^2 \psi/\partial x^2$
and $t=\partial^2 \psi/\partial u^2$,  
Eqs. (\ref{31}) and (\ref{32})  are written
in the respective forms
\begin{equation}
4xp^2 +\frac{1}{x h^2}q^2 = 2(i +\frac{1}{h^2}j +\frac{1}{h^4}k)
							  \label{33}
\end{equation}
and
\begin{equation}
\frac{4x}{h^2}r + 4l^2p +\frac{1}{xh^2}t  = - w -\frac{1}{h^2}f
		- \frac{1}{h^4}g -\frac{1}{h^6} d.
						\label{34}
\end{equation}
To integrate Eqs. (\ref{33}) and (\ref{34}) we apply
 a procedure suggested by Palumbo \cite{Pa}.
  Accordingly, employing $R$ and $\psi$ as independent coordinates 
instead of $R$ and $u $  (then $u=u(x, \psi))$, 
 we have
\begin{equation}
r=\left.\frac{\partial p}{\partial x}\right|_u=
  \left.\frac{\partial p}{\partial x}\right|_\psi
  +p\left. \frac{\partial p}{\partial\psi}\right|_x    
						\label{35}
\end{equation}
and
\begin{equation}
t=\left.\frac{\partial q}{\partial u}\right|_x
   = q\left.\frac{\partial q}{\partial \psi}\right|_x.
						\label{36}
\end{equation}
With the aid of Eqs. (\ref{33}), (\ref{35}),
(\ref{36}) and $f+i^\prime=0$ (which is satisfied identically),
Eq. (\ref{34}) reduces to 
\begin{equation}
\frac{4x}{h^2}\left .\frac{\pars p}{\pars x}\right|_{\psi} 
+ 4l^2p   = - w -\frac{1}{h^2}(f +i^\prime)
		-\frac{1}{h^4}(g+j^\prime) -\frac{1}{h^6} (d+ m^\prime).
						\label{37}
\end{equation}
Eq. (\ref{37}) can be considered  as an ordinary  linear first-order  equation
for $p(x,\psi )$ with respect to $x$ and constant $\psi$. Its solution
is
\beq
p=\frac{1}{4k^2x}(w + l^2 \tau(\psi))+\frac{\tau(\psi)}{4} - 
  \frac{l^2+k^2 x}{4}
  \left[(g+j^\prime) + l^2(d+m^\prime) + k^2(d+m^\prime)\frac{x}{2}\right],  
						     \label{38}
\eeq
where $\tau(\psi)$ is an integration  ``constant".

Since $u$ is a function of $x$ and $\psi$, solutions of equation
\begin{equation}
du = -\frac{p}{q} dx + \frac{1}{q} d \psi
						  \label{38a}   
\end{equation}
exist provided 
\begin{equation}
\frac{\partial}{\partial \psi}\left(-\frac{p}{q}\right)
 = \frac{\partial}{\partial x}\left(\frac{1}{q}\right).
						\label{39}
\end{equation}
Eq. (\ref{39})   leads to  
 the solvability condition
\begin{equation}
-q^2\frac{\partial p}{\partial \psi}
 + \frac{1}{2}p\frac{\partial q^2}{\partial \psi}
  + \frac{1}{2}\frac{\partial q^2}{\partial x} = 0.
						\label{40}
\end{equation}
Substituting   $q^2$ and $p$ from
 Eqs. (\ref{33}) and (\ref{38}) in Eq. (\ref{40}) yields  
a relation of the form  $\sum^7_{j=0} a_j(\psi)x^j=0$; 
hence, since $x$ and $\psi$ are independent variables, 
the equations $a_j=0$ ($j=0,\ldots, 7$) 
should be satisfied  for all $j$.   Consequently, the equilibrium problem  
is reduced to a set of eight ordinary non-linear differential equations 
for $\psi$
containing the  surface functions $P(\psi)$, $P_s(\psi)$, $F(\psi)$, 
$\Phi(\psi)$, $X(\psi)$, $\rho(\psi)$ and $\tau(\psi)$. It is therefore 
over-determined. 
Furthermore, to completely solve the equilibrium
problem, one should have to obtain the function $u(x,\psi)$ 
which by Eq. (\ref{38a}) satisfies the  
relation
\begin{equation}
\left.\frac{\partial  u}{\partial  x}\right|_\psi =
-\frac{p(x,\psi)}{q(x,\psi)} 
						   \label{41}
\end{equation}
where  $p(x, \psi)$ is given by Eq. (\ref{38}) and 
\begin{equation}
q(x,\psi) =xh^2 \left[2(i +\frac{1}{h^2}j + \frac{1}{h^4}m) 
	   -4x p^2\right]
						    \label{42}
\end{equation}
by Eq. (\ref{33}). Therefore, the existence of $T=T(\psi)$ helically symmetric
equilibria with incompressible flows, remains an open question.
It may be noted that from the present and our previous studies 
\cite{TaTh98,ThTa97} 
it  turns out that, 
as concerns the  mathematical structure and the physical properties  
of $T=T(\psi)$ equilibria with incompressible flows, there are 
similarities with 
static omnigenous 
equilibria.
The non-existence of {\em three dimensional} static omnigenous equilibria 
has been proved in an elegant manner by Palumbo and Balzano \cite{Pa86,PaBa}.
In the presence of flow, however, there are difficulties in applying 
their method because of the departure of the isobaric and current density
surfaces from the magnetic surfaces.

\begin{center}
{\large\bf IV.\ \ Conclusions}
\end{center}

The equilibrium of  helically symmetric plasmas with incompressible 
flows has been investigated within the framework of ideal magnetohydrodynamic
theory.
For the system under consideration the equilibrium equations
reduce 
to  a  second-order elliptic
partial differential equation for the helical magnetic flux function $\psi$
[Eq. (\ref{21a})] containing
the density $\rho(\psi)$, the electrostatic
potential $\Phi(\psi)$, the static equilibrium pressure $P_s(\psi)$, 
the function $F(\psi)$  associated with the helical flow 
and the function  $X(\psi)$  related to the  magnetic field component
along the symmetry direction 
in conjunction with  a relation for the pressure [Eq. (\ref{19})].
This equation has been  transformed to one,   possessing differential
part identical to the corresponding  static equilibrium equation 
[Eq. (\ref{23})], 
which permits the construction of several classes
of non-constant Mach-number analytic solutions. We have obtained 
exact solutions with flows in the direction of symmetry, 
  parallel to 
the magnetic  field  and  in arbitrary direction 
with differentially  varying electric fields 
perpendicular to the magnetic surfaces. 

For the physically appealing case of equilibria with isothermal
magnetic surfaces it has been shown that the equilibrium equations 
become over-determined, viz. the equilibrium problem reduces to
a set of eight ordinary differential equations for
$\psi$  with seven surface quantities. This  result is different from
the corresponding ones of cylindrical and axisymmetric 
equilibria for which the condition $T(\psi)$    imposes    only
 restrictions to the shapes of  the magnetic surfaces.  
Thus, the existence of helically symmetric equilibria with isothermal
magnetic surfaces as well as 
of generic  three dimensional equilibria of this kind remain open questions.

\begin{center}
 {\large\bf Acknowledgments}
\end{center}

Part of this work was conducted during a visit by one of the authors 
(G.N.T.) to  the Max-Planck Institut  f\"ur Plasmaphysik, Garching.
The hospitality of that Institute is greatly appreciated.

G.N.T. acknowledges support by EURATOM (Mobility Contract No
131-83-7 FUSC).
\newpage
\end{document}